\documentclass[aps,prl,showpacs,twocolumn]{revtex4}
\usepackage{graphicx}
\usepackage{bm}
\usepackage{amsmath}
\usepackage{amsfonts}

\def\ra{\rangle}
\def\la{\langle}

\def\dag{^\dagger}

\begin{document}

\preprint{}

\title{Electro-spinon in one-dimensional Mott insulator}
\author{
Hosho Katsura$^{1}$, Masahiro Sato$^{2}$, Takashi Furuta$^{3}$, and 
Naoto Nagaosa$^{1,3}$}
\affiliation{%
$^1$Cross-Correlated Materials T. Research Group (CMRG), RIKEN, Saitama,
Wako 351-0198, Japan\\
$^2$Condensed Matter Theory Laboratory, RIKEN, Saitama, Wako 351-0198,
Japan\\ 
$^3$Department of Applied Physics, The University of Tokyo, Tokyo
113-8656, Japan}%
\date{}
\begin{abstract}  
The low-energy dynamical optical response 
of 
dimerized and undimerized spin liquid states 
in a one-dimensional charge transfer 
Mott insulator is theoretically studied. 
An exact 
analysis is given for the low-energy asymptotic behavior 
using conformal field theory for the undimerized state.
In the dimerized state, the infrared absorption due to the bound state of 
two solitons, i.e, the breather mode, is predicted with an accurate 
estimate for its oscillator strength, 
offering a way to detect experimentally the excited singlet state. 
Effects of external magnetic fields are also discussed.    
\end{abstract}


\maketitle
Recently, the optical activity of 
insulating magnets 
has been attracting intensive interests 
\cite{Lorenzana_Sawatzky, Loidl_g, Loidl_g2, Kida}. 
Usually, the Mott insulator is considered to be inert 
in the lower energy region compared with the Mott gap. 
However, there are rich structures of the 
dielectric response of the insulating magnets in the infrared region as
revealed by recent experiments in multiferroics~\cite{Loidl_g, Loidl_g2, Kida}, which are explained 
partly by theories~\cite{KNB2, Aguilar_Miyahara}. 
The central concept there is the electro-magnon, i.e., 
the combined wave of spin and polarization, which can be detected both 
by neutron scattering experiment and optical absorption. 
There are basically two mechanisms for the 
optical activity due to the electro-magnons.  
One is due to the spin-orbit interaction~\cite{KNB2}, which is believed to
be the origin of the multiferroic behavior in 
$R$MnO$_3$ ($R$=Tb, Dy)~\cite{Kimura_Tb}. The electro-magnon absorption
has been first observed in these materials~\cite{Loidl_g}. 
Although the spin current model based on the spin-orbit interaction~\cite{KNB1} 
explains well the static polarization in $R$MnO$_3$, 
it turned out that this mechanism 
gives too weak oscillator strength for the infrared absorption.
The other mechanism is the exchange striction where the 
polarization is given by 
${\vec P} =  \sum_{ij}{\vec \Pi}_{ij} {\vec S}_i \cdot {\vec S}_j$
in the absence of inversion symmetry at the center of the bond connecting the sites
$i$ and $j$~\cite{Tanabe_Moriya_Sugano}.
This mechanism does not require the spin-orbit interaction, but 
usually gives rise to the two-magnon absorption.
However, the recent finding is that it leads to the single-magnon absorption
in non-collinear spin structures~\cite{Aguilar_Miyahara}. 
This can be easily seen 
by writing ${\vec S}_i = \la {\vec S}_i \ra + \delta {\vec S}_i$,
and consider the contribution to ${\vec P}$ linear in the fluctuation
$\delta {\vec S}_i$ from the ordered moment $\la {\vec S}_i \ra$.
The linear terms cancels when $\la {\vec S}_i \ra$ and $\la {\vec S}_j \ra$
are collinear while they survive for the noncollinear structure.

In this paper, we study theoretically the optical 
conductivity $\sigma(\omega)$ 
of the spin liquid state where the spin order is lost and the noncollinear 
spin structure is dynamically generated. For that purpose, we propose that
the (quasi) one-dimensional (1D) charge transfer Mott insulator is an
ideal arena where 
optical spectra provide a unique
information on the quantum dynamics of the spins. 
In 1D spin liquids, spinon 
(spin-$\frac{1}{2}$ object) 
is the most fundamental
particle~\cite{Gogolin_Nersesyan_Tsvelik} and the {\it electro-spinon}
governs the infrared activity. 
Especially, in the undimerized state, 
the low-energy asymptotic behavior of $\sigma(\omega)$ can be
analyzed by conformal field theory (CFT).
Furthermore, in the dimerized state, it is shown that the magnetically silent singlet
excitations can be detected by optical means with a precise prediction of 
the oscillator strength.

In the charge transfer compounds, the alternating stack of 
donor ($D$) and acceptor ($A$) molecules forms a chain. 
The simplified model for this system is the half-filled ionic 
Hubbard model~\cite{Nagaosa_Takimoto1}:
\begin{eqnarray}
H &=& - \sum_{i \sigma} t_i ( c^\dagger_{i \sigma} c_{i+1 \sigma} + {\rm h.c.})
+ 
\Big[ (-1)^i \frac{\Delta}{2}  - e \phi_i \Big] 
n_{i \sigma}
\nonumber \\
&+& U \sum_i n_{i \uparrow} n_{i \downarrow} + \frac{\kappa}{2} \sum_i u_i^2,
\end{eqnarray}
where $\Delta$ is the difference between the
energy levels of 
$D$ and $A$, 
$t_i =t_0 + g^\prime u_i$ is the transfer integral modulated by the molecular
displacement $u_i$ through the coupling constant $g^\prime$, 
and $\phi_i$ is the electric potential at the molecular site $i$.
The other notations are standard.
Neglecting $t_i$, the ground state is the ionic Mott insulator 
with one electron on each site when $U > \Delta$. On the other hand,  
the ground state is neutral for $U < \Delta$, i.e., all the odd sites are
fully occupied while the even ones are empty. 
\begin{figure*}[tb]
\begin{center}
\hspace{-.0cm}\includegraphics[width=1.9\columnwidth]{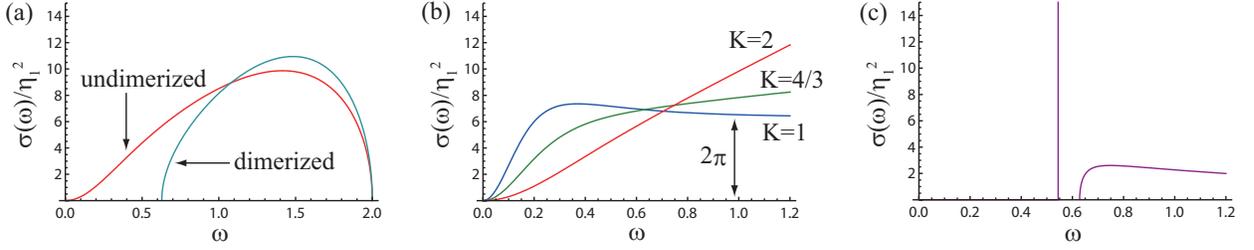}
\caption{
Optical conductivity $\sigma(\omega)$ of 1D Mott insulators. We take the unit $J={\cal C}=a=1$. 
(a) $\sigma(\omega)$ of the XY spin chain for
 $g u_0=0.31$ (dimerized) and that for $g u_0 =0$
 (undimerized; $K=2$) at $T=\frac{1}{4\pi}$. Note that $\eta=\pi \eta_1/\sqrt{2}$ when $K=2$. 
(b) Low-energy behavior of $\sigma(\omega)$ of the undimerized XXZ chain for $K=1, \frac{4}{3}$, and $2$ at $T=\frac{1}{4\pi}$, which is obtained from Eq. (\ref{eq:green}) by putting $q=0$. The spinon velocity is given by $v=\pi Ja/2=\pi/2$ when $K=1$. 
Due to the interaction effect, $\sigma(\omega)$ is enhanced at low energies as $K$ is decreased. 
Note that CFT description is valid only if $\omega, T \ll J=1$. 
In the limit of $\omega \gg T$, $\sigma(\omega)$ for $K=1$ approaches to the constant value $2\pi \eta^2_1$.  
(c) Low-energy behavior of $\sigma(\omega)$ of the dimerized Heisenberg chain with $M=0.2$ and $v=\pi/2$ at $T=0$. 
Delta function peak is located at $\omega=\sqrt{3} M v$. 
The $s$-${\bar s}$ continuum starts at $\omega=2Mv$. 
Comparing (b) and (c), the continuum intensity is transferred to the breather peak.
}
\label{fig:i-iii}
\end{center}
\end{figure*}

We are interested in the ionic state, and its 
effective spin Hamiltonian 
up to the linear order in $u$ and $\phi$ is given as follows: 
\begin{equation}
H_{\rm eff} = \sum_i 
[J +  g u_i  + \eta (-1)^i ( \phi_{i+1} - \phi_i ) ]
{\vec S}_i \cdot {\vec S}_{i+1} 
\label{eq:sp}
\end{equation}
with 
$J = 4 t_0^2 U/( U^2 - \Delta^2)$, 
$g = 8g^\prime t_0 U/(U^2 - \Delta^2)$, and 
$\eta= 8 e t^2_0 U \Delta/ (U^2 - \Delta^2)^2$. 
Because of the lack of inversion symmetry at the center of the
bond connecting $i$ and $i+1$, we have the polarization 
along the chain, which is given by the derivative of
$H_{\rm eff}$ with respect to the electric field: 
\begin{equation}
P = -\frac{\partial H_{\rm eff}}{\partial E} = \eta a \sum_{i} (-1)^{i} {\vec S}_i \cdot {\vec S}_{i+1}, 
\label{eq:pol}
\end{equation}
where $a$ is the lattice spacing. 
In real materials such as TTF-CA and TTF-BA~\cite{Torrance, Tokura_TTF},
the coupling to the molecular displacement $u_i$ leads
to the dimerization $u_i = u_0 (-1)^i$
accompanied by the 
singlet dimer state, i.e., the spin-Peierls instability occurs below a
transition temperature $T_{\rm SP}$. 
In this case, the uniform ferroelectric moment appears according to
Eq. (\ref{eq:pol}). 


Let us now study the optical spectra in this system. 
From the Kubo formula, the optical conductivity $\sigma(\omega)$ is given by 
\begin{equation}
\sigma(\omega) = 
\frac{\cal C}{La} \omega\,
{\rm Re} \int_0^\infty dt e^{(i \omega-\delta) t} \la [ P(t), P(0)] \ra,
\label{eq:Kubo_formula}
\end{equation}
where $L$ is the total number of sites in the chain and ${\cal C}=a/(\hbar V_{\rm u.c.})$
with the volume of the unit cell $V_{\rm u.c.}$. 
Using the 
Jordan-Wigner transformation, the effective spin model in the absence of
external fields is written as
\begin{equation}
H_{\rm eff} = \sum_i 
{{J^{\perp}_i} \over 2} 
( f_i^\dagger f_{i+1} + {\rm h.c.} ) + J^{z}_i \Big( n_i-\frac{1}{2} \Big)
\Big( n_{i+1}-\frac{1}{2}\Big),
\label{eq: h_eff}
\end{equation}
where $J^\alpha_i =J^\alpha (1+(-1)^i g u_0 /J)$ for 
$\alpha=\perp$ or $z$ with $J^\perp=J^z=J$.
Here $f_i$ and $f^\dagger_i$ denote the spinless fermions, 
and $n_i=f^\dagger_i f_i$. 
The XY model with $J^z=0$ is equivalent to the free-fermion
model. It can be analyzed exactly and gives a useful starting point to
analyze the realistic case $J^z=J^{\perp}$. 
Supposing that the polarization operator for $J^z=0$ is given by
$P=\eta a \sum_i (-1)^i (S^x_i S^x_{i+1}+S^y_i S^y_{i+1})$, 
the optical conductivity $\sigma(\omega)$ is evaluated as
\begin{equation}
\hspace{-2mm}\sigma(\omega) = {\cal D} \sqrt{( \omega^2 - \omega_1^2) (\omega_2^2 -
 \omega^2)}\, [f (-\omega/2)-f(\omega/2)],
\end{equation}
where $\omega_1 = 2g u_0$ is the spin gap induced by the dimerization,
while $\omega_2 = 2J$ is the bandwidth of the spinon dispersion. 
The coefficient ${\cal D}$ is given by ${\cal D}=\eta^2 {\cal C}a/[J^2 (1-\omega^2_1/\omega^2_2)^2]$ and
the Fermi distribution function is denoted by $f(E)=(e^{E/T}+1)^{-1}$ with $k_{\rm B}=\hbar=1$. 
The behaviors of $\sigma(\omega)$ 
are shown in Fig. 1 (a).  
Note that in the undimerized state, $\sigma(\omega) \propto \omega^2/T$
for $\omega \ll T$. 


This result for the non-interacting fermions should be modified 
by the interaction term, i.e., $J^z$ terms in Eq. (\ref{eq: h_eff}). 
Physically, this term introduces the attractive interaction 
between the particle and hole, which leads to the {\it excitonic}
effect and enhances 
$\sigma(\omega)$ at lower energies.
In the weak-coupling limit, this many-body effect can be studied using 
field theoretical methods as described below and several theoretical predictions will be made.

It is well known that the low-energy physics of the undimerized Heisenberg chain
is described by a Tomonaga-Luttinger 
(TL) model with a boson field $\Phi(x)$, which is equivalent to 
$c=1$ CFT~\cite{Gogolin_Nersesyan_Tsvelik}. 
In addition, the polarization operator, i.e., the staggered dimerization
operator, can be bosonized as   
\begin{equation}
P = \eta \int dx\epsilon(x)= \eta_1 \int dx
\cos(\sqrt{2\pi}\Phi(x)),
\label{eq:bos_P}
\end{equation}
where 
$\epsilon(x)=(-1)^i \vec S_i \cdot \vec S_{i+1}$. 
Therefore, the continuum model for the dimerized Heisenberg chain Eq. (\ref{eq:sp}) 
is given by the sine-Gordon 
Hamiltonian~\cite{Essler_Tsvelik_Delfino}:
\begin{equation}
\hspace{-1mm}H_{\rm SG}=\frac{1}{2}\int dx \Big[K\Pi^2+\frac{1}{K}(\partial_x \Phi)^2
-4\mu \cos(\sqrt{2\pi}\Phi)\Big],
\label{eq:SG}
\end{equation}
where $\Pi(x)$ is 
the canonical conjugate to
$\Phi(x)$, which satisfies $[\Phi(x), \Pi(y)]=i \delta(x-y)$, 
the coupling constant $\mu\propto g u_0$, and $K$ is the TL parameter. 
Here we have set $a=1$ and the spinon velocity $v=1$. 
In the present notation, the SU(2)-symmetric point ($J^\perp=J^z$)
corresponds to $K=1$, while $K>1$ 
to the easy-plane anisotropic case ($J^\perp>|J^z|$).  
In the high temperature region above $T_{\rm SP}$, 
the lattice displacement $u_0$ is zero and hence we can set $\mu=0$ in
Eq. (\ref{eq:SG}). Then the model is reduced to the $c=1$ CFT. 
Since the conformal weight of 
$\epsilon(x)$ 
is $(\Delta, {\bar \Delta})=(K/4, K/4)$, the two-point correlation
function behaves as 
$\langle \epsilon(z) \epsilon(0) \rangle \sim 1/|z|^{K}$
with $z=\tau + ix$ where $\tau=it$ is the imaginary time. 
Using the finite-size scaling argument, its retarded Green's function 
at finite temperature $T$ is obtained as
\begin{equation}
G^{\rm R}(q, \omega)=-\sin \Big(\frac{\pi K}{2}\Big)(2\pi T)^{K-2}
F_K(q, \omega),
\label{eq:green}
\end{equation}
where 
$F_K(q,\omega)=B(\frac{K}{4}
-i\frac{\omega+q}{4\pi T},1-\frac{K}{2})
B(\frac{K}{4}-i\frac{\omega-q}{4\pi T},1-\frac{K}{2})$
with the Euler beta function $B(x,y)$~\cite{Schulz1}. From
Eqs. (\ref{eq:Kubo_formula}) and (\ref{eq:green}), we obtain $\sigma(\omega) = -\eta^2_1 {\cal C}\,\omega
{\rm Im} G^{\rm R}(0, \omega)$ for several values of $K$ as shown in Fig. 1 (b). 
The optical conductivity is actually enhanced at low energies with a decrease in $K$, 
i.e., the increase in $J_z$. 
Note that CFT description is valid only if $\omega$ and $T$ are much
smaller than $v/a \sim J$. 
For $\omega \ll T$, we find that $\sigma(\omega) \propto \omega^2 T^{K-3}$.
It is consistent with our earlier analysis of the XY model corresponding to $K=2$. 
On the other hand, 
we find $\sigma(\omega)\propto \omega^{K-1}$ for $\omega \gg T$. 
It is remarkable that $\sigma(\omega)$ is asymptotically constant at the
Heisenberg point ($K=1$). 
This constant is given by $2\pi \eta^2_1 {\cal C}a$. 
It is possible to estimate the unknown coefficient $\eta_1$ 
from the experimentally observed $\sigma(\omega)$.


In the low temperature region below 
$T_{\rm SP}$, $\mu$ in Eq. (\ref{eq:SG}) is nonzero and
has a significant effect on the system. 
Fortunately, the model is still exactly solvable
and one can apply powerful techniques such as the Bethe
ansatz and form factor expansion~\cite{Smirnov}. Let us concentrate on
the realistic $K=1$ case 
at $T=0$, where the spectrum of $H_{\rm SG}$ consists of four
quasi-particles, i.e., kink ($s$), anti-kink (${\bar s}$), and light
($b1$) and heavy ($b2$) breathers. 
The breathers are the bound states of the kinks and
anti-kinks. Due to the hidden SU(2) symmetry, $s$,
${\bar s}$, and $b1$ form a spin-triplet with the mass gap $M$ while
$b2$ is a spin-singlet with the mass $\sqrt{3}M$. 
The formula for the dimensionless gap $M$ can be found
in~\cite{Lukyanov_Zamolodchikov}:  
\begin{equation}
M=\frac{2\Gamma(1/6)}{\sqrt{\pi}\Gamma(2/3)} 
\left(\frac{\pi\Gamma(3/4)}{\Gamma(1/4)}\mu \right)^{2/3} 
\equiv{\cal A}\cdot (2\mu)^{2/3},
\label{eq:Mass}
\end{equation}
where ${\cal A} \approx 3.041$ and $\Gamma(x)$ denotes the Euler gamma function. 
Physically, it corresponds to the spin-Peierls gap divided by $\hbar v/a=\pi J/2$. 
Suppose that the relation between the polarization density and the bosonic field is given
by $p(x) = \eta_1 \cos(\sqrt{2\pi}\Phi(x))$. 
Then the spontaneous polarization can be expressed by $M$ as 
\begin{equation}
\langle p(x) \rangle = ({\cal A}/3)^{3/2} \eta_1 \sqrt{M}
\label{eq: sppol}
\end{equation}
using the vacuum
expectation value of the vertex operator 
$e^{i\beta \Phi}$~\cite{Lukyanov_Zamolodchikov}. Therefore, from the 
polarization measured in experiments, one can determine $\eta_1 \sqrt{M}$ exactly.

Next, we introduce the form factors and discuss the optical conductivity
at $T=0$. 
The Fock space of the sine-Gordon model can be constructed from the
creation operators for the quasi-particles:
$
|\theta_1, \theta_2 ... \theta_n
\ra_{\epsilon_1, \epsilon_2 ...\epsilon_n}
=A\dag_{\epsilon_1} (\theta_1) A\dag_{\epsilon_2} (\theta_2) 
... A\dag_{\epsilon_n} (\theta_n)|0\ra.
$
Here $A\dag_{\epsilon_k}(\theta_k)$, the creation operator for 
$\epsilon_k=s,\, {\bar s},\, b1$, or $b2$ with the mass $M_{\epsilon_k}$,
depends on the rapidity variable $\theta_k$ which is related to the energy
and momentum through the relations: $E_k=M_{\epsilon_k} \cosh \theta_k$
and $P_k =M_{\epsilon_k} \sinh \theta_k$. 
The operators $A\dag_{\epsilon_i}(\theta_i)$ and
$A_{\epsilon_j} (\theta_j)$ are neither bosons nor fermions and satisfy
nontrivial commutation relations~\cite{Gogolin_Nersesyan_Tsvelik}.
In this Fock space, the form factor of $\cos(\sqrt{2\pi}\Phi)$ is defined as
\begin{equation}
F^{\cos}(\theta_1 ... \theta_n)_{\epsilon_1 ... \epsilon_n} 
= \la 0 | \cos (\sqrt{2\pi}\Phi (0, 0))
|\theta_1 ... \theta_n\ra_{\epsilon_1 ... \epsilon_n}. \nonumber 
\label{eq:form}
\end{equation}
The optical conductivity can be expressed in terms of the form factors as
\begin{eqnarray}
\sigma(\omega) &=& \displaystyle 2\pi^2 \eta^2_1 {\cal C}\, \omega 
\sum^{\infty}_{n=1} \sum_{\epsilon_i} 
\int \frac{d\theta_1 \cdots \theta_n}{(2\pi)^n n!} 
|F^{\cos}(\theta_1 ... \theta_n)_{\epsilon_1 ... \epsilon_n}|^2
\nonumber \\
& \times & \textstyle{\delta(\sum_j P_j)\delta(\omega-\sum_j E_j)
}. 
\label{eq: formal form}
\end{eqnarray}
To study the above expansion, it is useful to note that $H_{\rm SG}$ is
invariant under the charge conjugation, $C \Phi C^{-1}=-\Phi$.
From this symmetry and $C A\dag_{b2} C^{-1}=A\dag_{b2}$, the first
nontrivial contribution stems from the heavy-breather form
factor~\cite{Essler_Tsvelik_Delfino, Lukyanov, Demler, Kuzmenko_Essler}. The
absolute square of this form factor is independent of the rapidity
$\theta$ and is given by
\begin{eqnarray}
|F^{\cos}_{b2}|^2 &=& \frac{{\cal A}^3 M}{2\cdot 3^{3/2}} 
\exp\left[
-4\int^\infty_0 \frac{dx}{x} \frac{\cosh(\frac{\pi x}{6}) 
\sinh^2(\frac{\pi x}{3})}{\sinh(\pi x) \cosh(\frac{\pi x}{2})}
\right] \nonumber \\
& \approx & 1.077 M.
\label{eq: breather form}
\end{eqnarray}
From Eqs. (\ref{eq: formal form}) and (\ref{eq: breather form}), the optical conductivity due to the heavy breather is 
exactly given by
\begin{equation}
\sigma_{b2}(\omega)=\pi \eta^2_1 {\cal C}\, \omega 
\frac{|F^{\cos}_{b2}|^2}{\sqrt{3}M} \delta(\omega-\sqrt{3}M). 
\label{eq:breather}
\end{equation}
The next-leading contribution comes from the $s$-${\bar s}$
continuum (see Fig. 1 (c)). The threshold of this continuum is at $\omega=2M$. 
Therefore, the integrated intensity up to this
threshold is given by
\begin{equation}
\textstyle{
\int^{2M}_0 d\omega \sigma_{b2}(\omega) = 
\pi \eta^2_1 {\cal C} v |F^{\cos}_{b2}|^2 \propto \eta^2_1 M
}
\end{equation} 
with the insertion of $v$ and $a$. 
Combining this oscillator strength with the peak position located at
$\omega=\sqrt{3}M$, 
one can {\it exactly} determine the coefficient $\eta_1$ experimentally. 
At the same time, from the spontaneous polarization Eq. (\ref{eq: sppol}), 
one can also determine $\eta_1$ and compare it with that obtained
from the oscillator strength. 
Therefore, the validity of the sine-Gordon theory can be tested by 
means of optical conductivity and spontaneous polarization measurements.  
Let us now estimate the actual value of the integrated intensity for TTF-BA. 
This compound shows the spin-Peierls transition below $T_{\rm SP}=53$ K and 
the spontaneous polarization along $b$-axis is measured as 
$P_{\rm s} \sim 0.15 \mu$C/cm$^2$~\cite{Kagawa}. 
Combining with $J \sim 170$K~\cite{Girlando_Pecile} and 
$V_{\rm u.c.} \sim 8.4 \times 4.3 \times 11.6~\AA^3$~\cite{Garcia},
we estimate the oscillator strength as 
$\int^{2M}_0 d\omega \sigma_{b2}(\omega) \sim 10^{13}~\Omega^{-1}$cm$^{-1}$s$^{-1}$, 
which can be measured by infrared absorption spectroscopy. 


\begin{figure}[tb]
\begin{center}
\hspace{-.0cm}\includegraphics[width=0.8\columnwidth]{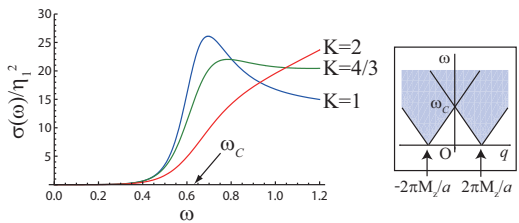}
\caption{Optical conductivity $\sigma(\omega)$ of the field-induced TL 
liquid with magnetization $M_z=0.1$ for $K=1, \frac{4}{3}$, and $2$ at $T=\frac{1}{8\pi}$. We use the unit $v=a=1$. The location of $\omega_c=2 \pi M_z v/a$ is indicated. 
The imaginary part of ${\tilde G}^{\rm R}(q, \omega)$ is nonzero in the shaded region of the right figure. 
Note again that CFT description is valid only if $\omega, T \ll v=1$. 
}
\label{fig:mag_field}
\end{center}
\end{figure}

Next, we shall consider the effect of 
the Zeeman coupling term  
$H_{\rm mag}=-g\mu_B H \sum_i S^z_i$. 
As long as the magnetic field $h=g\mu_BH$ is smaller than the spin-Peierls gap $M$, 
the singlet excitation ($b2$) is insensitive to the field while the energies of the triplet excitations 
($s$, $\bar s$, and $b1$) are split.  
On the other hand, if $h>M$, the one species of the triplet particles 
becomes gapless and a uniform magnetization $M_z=\langle S_i^z\rangle$ appears. 
Its effective Hamiltonian is equal to a TL model with a new boson field 
$\tilde\Phi(x)=\Phi(x)-\sqrt{2\pi}M_z x$. 
The polarization density can be rewritten as 
$
\tilde p(x)=\eta_1\cos(\sqrt{2\pi}\tilde\Phi(x)-2\pi M_z x).
$
Note that this TL-liquid state is adiabatically connected to that of 
the undimerized Heisenberg chain ($u_0=0$) under the magnetic field. 
The magnetization dependence of the TL parameter $K$ in dimerized and undimerized Heisenberg chains can be found in~\cite{Sakai} and~\cite{Cabra}, respectively. 

Applying CFT, we can again evaluate the correlation function 
of the polarization as in Eq.~(\ref{eq:green}). The two-point correlator 
behaves as $\langle \tilde p(z)\tilde p(0)\rangle\sim \cos(2\pi M_z x)/|z|^K$, 
and the retarded Green's function around the wavenumber $q=0$ 
is represented as ${\tilde G}^{\rm R}(q,\omega)=G^{\rm R}(q-2\pi M_z, \omega)+G^{\rm R}(q+2\pi M_z,\omega)$. 
From this, we can obtain 
$\sigma(\omega) = -\eta^2_1 {\cal C}\, \omega{\rm Im}{\tilde G}^{\rm R}(0,\omega)$ as shown in Fig.~\ref{fig:mag_field}. 
Remarkably, in sufficiently low temperatures $T\ll v/a$, $\sigma(\omega)$ 
starts from a finite frequency around $\omega_c = 2\pi M_z v/a$. 
This is in sharp contrast to the gapless behavior in the zero-magnetization case
[see Fig.~\ref{fig:i-iii} (b)]. 


Lastly, we compare the above results with those of Raman
scattering~\cite{Raman}. 
The Raman spectrum is proportional to the dynamical structure factor of
$R\propto \sum_i\vec S_i\cdot\vec S_{i+1}$ for the spin chain Eq. (\ref{eq:sp}). For the dimerized case, the Hamiltonian does not commute with $R$ and hence its spectrum is equivalent to the dynamical structure factor of ${\tilde R} \propto \sum_i (-1)^i \vec S_i\cdot\vec S_{i+1}$. 
Therefore, 
the Raman scattering can also provide a spectrum with a singlet peak
as has indeed been observed in CuGeO$_3$~\cite{CuGeO3}. 
However, we should note that it is difficult to theoretically
predict the intensity of the Raman spectrum, in general, while 
we can evaluate that of the absorption with considerable accuracy 
as in Eq.~(\ref{eq:breather}). Furthermore, the infrared absorption can detect 
gapless spinon excitations in the undimerized case  
where any contribution from spinons is absent in 
the Raman scattering since the operator $R$ commutes with $H_{\rm eff}$ with $u_i=0$. 


In summary, we have studied the optical activity of the electro-spinons
in the undimerized and dimerized quantum spin chains with $S=\frac{1}{2}$
derived from the ionic Hubbard model. The scaling form of the low-energy
$\sigma(\omega)$ is predicted by applying the conformal field theory in
the undimerized case, and the exact estimate of the oscillator strength
due to the singlet breather mode is given. These theoretical predictions can be
tested in organic charge transfer Mott insulators.
   

The authors are grateful to J. Fujioka, F. Kagawa, Y. Tokura, and A. M. Tsvelik for useful discussions. 
This work is supported in part by Grant-in-Aids 
(Grant No.~15104006, No.~16076205, No.~17105002, No.~19048015, No.~19048008, No.~21244053, No.~21740295) 
and NAREGI Nanoscience Project from MEXT. 


\begin{thebibliography}{99}
\bibitem{Lorenzana_Sawatzky}
J. Lorenzana and G. A. Sawatzky, Phys. Rev. B \textbf{52}, 9576 (1995).

\bibitem{Loidl_g}
A. Pimenov {\it et al.}, Nature Phys. \textbf{2}, 97 (2006).

\bibitem{Loidl_g2}
A. Pimenov {\it et al.}, Phys. Rev. B \textbf{74}, 100403(R) (2006). 

\bibitem{Kida}
N. Kida {\it et al.}, Phys. Rev. B \textbf{78}, 104414 (2008).

\bibitem{KNB2}
H. Katsura, A. V. Balatsky, and N. Nagaosa, Phys. Rev. Lett. \textbf{98}, 027203 (2007).

\bibitem{Aguilar_Miyahara}
R. Vald${\acute {\rm e}}$s Aguilar {\it et al.}, Phys. Rev. Lett. \textbf{102}, 047203 (2009); 
S. Miyahara and N. Furukawa, 	arXiv:0811.4082.

\bibitem{Kimura_Tb}
T. Kimura, {\it et al.}, Nature \textbf{426}, 55 (2003); T. Goto {\it et al.}, Phys. Rev. Lett. \textbf{92}, 257201 (2004). 

\bibitem{KNB1}
H. Katsura, N. Nagaosa, and A. V. Balatsky, Phys. Rev. Lett. \textbf{95}, 057205 (2005). 


\bibitem{Tanabe_Moriya_Sugano}
Y. Tanabe, T. Moriya, and S. Sugano, Phys. Rev. Lett. \textbf{15}, 1023 (1965).

\bibitem{Gogolin_Nersesyan_Tsvelik}
A. O. Gogolin, A. A. Nersesyan, and A. M. Tsvelik, {\it Bosonization and Strongly Correlated Systems} (Cambridge University Press, Cambridge, 1998).

\bibitem{Nagaosa_Takimoto1}
N. Nagaosa and J. Takimoto, J. Phys. Soc. Jpn, \textbf{55}, 2735 (1986); {\it ibid}., \textbf{55}, 2754 (1986).

\bibitem{Torrance}
J. B. Torrance {\it et al.}, Phys. Rev. Lett. \textbf{46}, 253 (1981);  {\it ibid}., \textbf{47}, 1747 (1981). 

\bibitem{Tokura_TTF}
Y. Tokura {\it et al.}, Phys. Rev. Lett. \textbf{63}, 2405 (1989).


\bibitem{Essler_Tsvelik_Delfino}
F. H. L. Essler, A. M. Tsvelik, and G. Delfino, Phys. Rev. B \textbf{56}, 11001 (1997).

\bibitem{Schulz1}
H. J. Schulz, Phys. Rev. B \textbf{34}, 6372 (1986). 

\bibitem{Smirnov}
F. A. Smirnov, {\it Form Factors in Completely Integrable Models of Quantum Field Theory} (World Scientific, Singapore, 1992).

\bibitem{Lukyanov_Zamolodchikov}
S. Lukyanov and A. Zamolodchikov, Nucl. Phys. B \textbf{493}, 571 (1997). 
\bibitem{Lukyanov}
S. Lukyanov, Mod. Phys. Lett. A \textbf{12}, 2543 (1997). 

\bibitem{Demler}
V. Gritsev, A. Polkovnikov, and E. Demler, Phys. Rev. B 75, 174511 (2007). 
\bibitem{Kuzmenko_Essler}
I. Kuzmenko and F. H. L. Essler, Phys. Rev. B \textbf{79}, 024402 (2009). 

\bibitem{Kagawa}
F. Kagawa {\it et al.}, unpublished. 
\bibitem{Girlando_Pecile}
A. Girlando and C. Pecile, Solid. State. Comm. \textbf{54}, 753 (1985).  
\bibitem{Garcia}
P. Garc$\acute{\i}$a {\it et al.}, Phys. Rev. B \textbf{72}, 104115 (2005).

\bibitem{Sakai}
T. Sakai, J. Phys. Soc. Jpn. {\bf 64}, 251 (1995);
T. Suzuki and S. I. Suga, Phys. Rev. B {\bf 70}, 054419 (2004).

\bibitem{Cabra}
D.C. Cabra, A. Honecker, and P. Pujol, Phys. Rev. B \textbf{58}, 
6241 (1998).

\bibitem{Raman}
See, for example, 
P. Lemmens, C. Gros and G. G\"ontherodt. Phys. Rep. {\bf 375}, 1 (2003).

\bibitem{CuGeO3}
V. N. Muthukumar, {\it et al.}, Phys. Rev. B {\bf 54}, R9635 (1996).
\end{thebibliography}
\end{document}